\documentclass[useAMS,usenatbib]{mn2e}
\usepackage[figuresright]{rotating}
\usepackage{rotating}
\usepackage{color}
\usepackage{longtable}
\usepackage{lscape}
\usepackage{lipsum}
\usepackage{lscape}
\usepackage{xspace}

\title[Chemical abundances of  Seyfert 2s]{Chemical abundances of  Seyfert 2 AGNs-- II.   
 $N2$ metallicity calibration  based on SDSS}

\author[Carvalho et al.]
  {S.~P. Carvalho$^{1}$, O.~L. Dors$^{1}$\thanks{E-mail: olidors@univap.br},
  M.~V. Cardaci$^{2}$, G.~F. H\"agele$^{2}$, A.~C. Krabbe$^{1}$, \newauthor{E. P{\'e}rez-Montero$^3$, 
  A.~F. Monteiro$^{1}$, M. Armah$^{1}$, P. Freitas-Lemes$^{1}$}\\
$^1$ UNIVAP - Universidade do Vale do Para{\'i}ba. Av. Shishima Hifumi, 2911, CEP:
12244-000, S{\~ a}o Jos{\'e} dos Campos, SP, Brazil\\
$^2$ Instituto de Astrof\'{\i}sica de La Plata (CONICET-UNLP), Argentina\\
$^3$ Instituto de Astrof{\'i}sica de Andaluc{\'i}a (CSIC), Camino Bajo de Hu{\'e}tor s/n, Aptdo. 3004, E18080-Granada, Spain.\\
}

\date{Released 2019 Apr 15}

\def\cm2{cm$^2$ }
\def\se1{s$^{-1}$ }

\newcommand{\ion}[2]{{\textrm{#1}}\,{\textrm{\sc #2}}}

\begin{document}

%\label{firstpage}

\maketitle

\begin{abstract}
We present a semi-empirical calibration between the metallicity ($Z$) of Seyfert 2 Active Galactic
Nuclei and the $N2$=log([\ion{N}{ii}]$\lambda$6584/H$\alpha$) emission-line intensity ratio.
This calibration was derived through the [\ion{O}{iii}]$\lambda$5007/[\ion{O}{ii}]$\lambda$3727
versus $N2$ diagram containing observational data and photoionization
model results obtained with the {\sc Cloudy} code. The observational sample consists of 
 463 confirmed Seyfert 2 nuclei (redshift $z \: \la 0.4$) taken from the Sloan Digital Sky Survey DR7 dataset.
The obtained $Z$-$N2$ relation  is valid for the range  $0.3 \: \la \: (Z/Z_{\odot}) \: \la \: 2.0$  which
corresponds to $-0.7 \: \la \: (N2) \: \la \: 0.6$.  The effects of varying
the  ionization parameter ($U$), electron density and the slope of the spectral energy distribution
on the $Z$ estimations are  of the order of the uncertainty produced by the error measurements of $N2$.
This result indicates the large reliability of our $Z-N2$ calibration.
A relation between  $U$ and the  [\ion{O}{iii}]/[\ion{O}{ii}] line ratio, almost independent
of other nebular parameter, was obtained.   
\end{abstract}

%________________________________________________________________________

\section{Introduction}
\label{intro}

Active Galactic Nuclei (AGNs) are the most luminous 
objects  in the Universe and present strong emission-lines
in their spectra. The metallicity derived through these
emission-lines offers a  very powerful tool for understanding 
the chemical galaxy evolution along the Hubble time.

 Among the heavy elements present in the gas phase of  gaseous nebulae,
oxygen is the element most widely used as a proxy for global gas-phase metallicity $Z$
(e.g. \citealt{kennicutt03, hagele08, yates12}) because prominent emission lines from their main ionic
stages are present in the optical spectra of these objects.
It is consensus that  {\it bona fide} oxygen abundance\footnote{The oxygen abundance is definied by  
the ratio of the number of  oxygen atoms to hydrogen atoms (O/H).} determinations in  
 star-forming regions and planetary nebulae are those based on direct detection 
of the electron temperature ($T_{\rm e}$) of the gas, the so-called $T_{\rm e}$-method. 
The agreement between
oxygen abundances in  the gas phase of \ion{H}{ii} regions 
with those derived through observations of the weak interstellar \ion{O}{i}$\lambda$1356  line towards the stars
located at similar galactocentric distance in the Milky Way \citep{pilyugin03} indicates   
the $T_{\rm e}$-method is consistent with other more 
precise ways of deriving  the metallicity. However, this method requires
the measurement of  certain weak emission-lines sensitive to $T_{\rm e}$, such as 
[\ion{O}{iii}]$\lambda$4363 ($\sim$100 times weaker than H$\beta$),
 which makes $T_{\rm e}$-method only applied to objects with high ionization degree  and/or low metallicity
(e.g. \citealt{smith75, castellanos02, kenniccutt03, izotov06, hagele08, sanders16, sanders19}, among others).
In the  cases where the $T_{\rm e}$-method can not be applied,  theoretical or 
(semi-) empirical calibrations between abundances or metallicity and more easily measurable line ratios 
can be used instead,  the so-called strong-line method (for a review, see \citealt{enrique17, peimbert17, kewley19, maiolino19, jorge20}).

 In regarding AGNs, the $T_{\rm e}$-method tends to underestimate the  
 oxygen abundance by an average value of about 0.6 dex in comparison  to estimations
based on strong-line methods and it produces  subsolar O/H values for   most of these objects
\citep{dors15, dors20}. An alternative method to derive the metallicity  or abundances in the nuclear regions of spiral galaxies
is the  extrapolation of the radial oxygen abundance.
Along decades, results based on this indirect method have indicated 
$Z$ near  or slightly above the solar value in  nuclear regions 
\citep{vilacostas92, zaritsky94, vanzee98, pilyugin04, gusev12, dors15, igor19},  
in consonance with predictions of chemical evolution models (e.g. \citealt{molla05})
and with the use of strong-line methods (e.g. \citealt{groves04, groves06, feltre16, thomas19, enrique19, dors20}).
Therefore,  $T_{\rm e}$-method does not seem to work for AGNs. 
The origin of the discrepancy between $Z$ values calculated via $T_{\rm e}$-method and via strong-line methods,
the so-called $T_{\rm e}$-problem, could  be attributed, in part, to the presence of  heating/ionization
by gas shock in the Narrow Line Region (NLR) of AGNs. In fact, \citet{contini17}  carried out  
detailed modelling of  AGN optical emission-lines  by using the SUMA code \citep{contini83} and  
suggested the presence of gas shock with low velocity ($v \: \la \: 400 \: \rm km \:s^{-1}$)
in  a sample of Seyfert 2 nuclei. This result is supported by recent spatially resolved observational studies of
Seyfert 2 nuclei, in which  the presence of gas outflows with velocity  of the order of 100-300 $\rm km \: s^{-1}$ 
have been found (e.g. \citealt{riffel17, riffel18}). Moreover, the $T_{\rm e}$-problem can also be  originated due to
the use of an unappropriate calculation of the Ionization Correction Factor (ICF) for oxygen in AGNs \citep{enrique19, dors20}.

The most common way to obtain a calibration between strong emission-lines and $Z$  (or O/H) is 
through the use of photoionization models. The basic idea is to calculate emission-line ratios  
sensitive to $Z$  taking into account their dependence  on
other nebular parameters such as, the ionization parameter ($U$) of the gas, electron density, among others. 
For the optical range, the first calibration based on photoionization models for
AGNs was proposed by \citet{thaisa98}, who used the line ratios [\ion{N}{ii}]$\lambda$$\lambda$6548,6594/H$\alpha$,
[\ion{O}{iii}]$\lambda$$\lambda$4949,5007/H$\beta$ and [\ion{O}{ii}]$\lambda$$\lambda$3726,3729/[\ion{O}{iii}]$\lambda$$\lambda$4949,5007.
In this case, [\ion{N}{ii}]/H$\alpha$ is the $Z$ indicator\footnote{\citet{thaisa98} assumed in their
calibrations $Z$=12+log(O/H).}, as proposed by \citet{thaisa94}, and the ratios involving 
[\ion{O}{iii}]  are mainly dependent on the ionization degree rather than $Z$.  
Most recently, \citet{castro17}, using a comparison between
photoionization models results and heterogeneous observational data of 58 Seyfert 2 nuclei,  proposed a semi-empirical calibration
of $Z$ with   the  $N2O2$=[\ion{N}{ii}]$\lambda$6584/[\ion{O}{ii}]$\lambda$3727 index.  Throughout the paper, 
[\ion{O}{ii}]$\lambda$3727  refers to the sum of  [\ion{O}{ii}]$\lambda$3726 and [\ion{O}{ii}]$\lambda$3729.

The   $N2O2$ line ratio  presents some advantages  over other $Z$ indicators.
Firstly, $N2O2$ is not bi-valued  as the most  widely used
 $R_{23}$=([\ion{O}{ii}]$\lambda$3727+[\ion{O}{iii}]$\lambda$$\lambda$4949,5007)/H$\beta$ index,
proposed by \citet{pagel79} and usually used in \ion{H}{ii} region studies. Thus, the $N2O2$   estimates
metallicities  in a wide range of $Z$ values ($0.5 \: \la \: (Z/Z_{\odot}) \: \la \: 2.0$; \citealt{castro17}). Secondly, $N2O2$
involves ions with similar ionization potentials, which minimizes the effects of the presence
of  possible secondary heating (ionizing) sources.
However, $N2O2$ suffers some limitations, mainly because it requires spectrophotometric data covering a
wide spectral range, making  the reddening correction  crucial. Moreover, in recent optical surveys, e.g.  
MaNGA (Mapping Nearby Galaxies at the Apache Point Observatory, \citealt{law15}), 
the [\ion{O}{ii}]$\lambda$3727 line  is measured in very few objects (e.g. \citealt{sandro17, janaina19}).
Even in the data from  the Sloan Digital Sky Survey (SDSS, \citealt{york00}),  
when the presence of the [\ion{O}{ii}]$\lambda$3727 line is considered in the selection criteria of objects,
the  sample is considerably reduced  (e.g. \citealt{pilyugin11}).
In this sense, the $N2$=log([\ion{N}{ii}]$\lambda$6584/H$\alpha$) seems to be a better $Z$ indicator than $N2O2$.

 In this paper,   the observational data of confirmed Seyfert~2 AGNs,  taken from
the Sloan Digital Sky Survey (SDSS, \citealt{york00})  DR7 and selected by \citet{dors20}, hereafter  
referred as Paper~I, were combined with photoionization model results in order to explore 
the feasibility of the [\ion{N}{ii}]$\lambda$6584/H$\alpha$ ratio as a metallicity indicator. This paper is organized  as follows: 
In Section~\ref{meth}, a   description  of the methodology used to obtain the $Z-N2$ calibration is presented. 
In Sect.~\ref{res}, a comparison between the observational data and  photoionization model results
as well as the calibration obtained are presented.
The discussion is presented in Sect.~\ref{disc}. In Sect.~\ref{conc}, 
the summary and the conclusions of the outcome are presented.

\section{Methodology}
\label{meth}

To obtain a calibration between the   $Z$ and the $N2$ index,  the same methodology 
used by \citet{castro17} and \citet{dors19} to calibrate the $Z$  with  optical and
ultraviolet NLRs associated to  type-2 AGNs, respectively, was adopted.
 Based on the [\ion{O}{iii}]$\lambda$5007/[\ion{O}{ii}]$\lambda$3727 
versus [\ion{N}{ii}]$\lambda$6584/H$\alpha$ diagram,
the observational data of Seyfert 2 AGNs  were compared  with photoionization model predictions. From this
diagram, for each object, the metallicity and the corresponding $N2$ value were obtained, resulting in an unidimensional
calibration. In what follows,  descriptions of the observational
sample and of the photoionization models are presented. 

\subsection{Observational data}
\label{obs}

We used optical emission-line intensities of Seyfert~2 nuclei taken from 
the Sloan Digital Sky Survey (SDSS, \citealt{abazajian09}) DR7   
 presented in Paper~I. These data comprehend reddening corrected  intensities (in relation to H$\beta$)  
of the   
[\ion{O}{ii}]$\lambda$3726+$\lambda$3729, 
[\ion{Ne}{iii}]$\lambda$3869, 
[\ion{O}{iii}]$\lambda$4363, 
%[\ion{O}{iii}]$\lambda$4959, 
[\ion{O}{iii}]$\lambda$5007, 
He\,I$\lambda$5876, 
[\ion{O}{i}]$\lambda$6300, 
%[\ion{N}{ii}]$\lambda$6548, 
H$\alpha$, 
[\ion{N}{ii}]$\lambda$6584, 
[\ion{S}{ii}]$\lambda$6716, 
[\ion{S}{ii}]$\lambda$6731 
and [\ion{Ar}{iii}]$\lambda$7135 emission-lines.
The line measurements were carried out by the MPA/JHU\footnote{Max-Planck-Institute for Astrophysics and John Hopkins University} group.

Observational data  taken from the SDSS have been widely used to derive
physical properties  of AGNs (e.g. \citealt{vaona12, zhang13}). However, in most cases,
the classification of AGN-like objects has been obtained by using only standard 
Baldwin-Phillips-Terlevich diagrams \citep{baldwin81, veilleux87},  which include, for instance,
Seyfert 1s, Seyfert 2s, quasars, \ion{H}{ii}-like objects with very strong winds and gas shocks. Therefore, with the
goal of selecting only Seyfert~2 objects, in the Paper~I, we used a set of diagnostic diagrams
 to select AGN-like objects. Subsequently, the resulting data sample were compared 
with their classification obtained from the NED/IPAC\footnote{ned.ipac.caltech.edu}
(NASA/IPAC Extragalactic Database) database in order to select only objects classified as Seyfert~2 nuclei.  This procedure resulted
in a sample of 463 Seyfert~2 nuclei with redshifts $z \: \la \: 0.4$ and with stellar masses of the hosting galaxies
(also taken from the MPA-JHU group) in the range of $9.4 \: \la \: \log(M/ \rm M_{\odot}) \: \la \: 11.6$. From the compiled sample,
we selected the intensities of the  [\ion{O}{ii}]$\lambda3727$, [\ion{O}{iii}]$\lambda5007$,
H$\alpha$, and [\ion{N}{ii}]$\lambda6584$ emission-lines relative to H$\beta$.  
The reader is referred to Paper~I for a complete description about the observational data  and
 aperture effects on $Z$ estimation.

\subsection{Photoionization models}
\label{mod}

We  considered  version 17.00 of the {\sc Cloudy} code \citep{ferland17} in order to build up
 photoionization model grids assuming a wide range of nebular parameters. These models are
similar to the ones considered by \citet{dors19}  and
the reader is referred to this paper for a complete description. 
The input parameters are described below.

\begin{enumerate}
\item SED: The Spectral Energy Distribution (SED)   was assumed to be composed  of the sum of two components:
one representing the Big Blue Bump peaking at $\rm 1 \: Ryd$, and  the other  a power law with spectral index $\alpha_x=-1$ 
 representing the non-thermal X-ray radiation. The continuum between 2 keV and 2500\AA\ is 
 described by a power law with a spectral index
 $\alpha_{ox}$, for which we consider three different values:  $-$0.8, $-$1.1 and $-1.4$, i.e. 
 about  the range of  values estimated for Seyfert 2
 and   Quasars (e.g. \citealt{ho99, miller11,zhu19}). It must be noted that models assuming $\alpha_{ox} \: < \: -1.4$
 predict very low emission-line intensities (relative to H$\beta$), when compared  to those from our observational data (see also \citealt{dors12}). 
Moreover, observational estimations of $\alpha_{ox}$ have shown that few  AGNs present $\alpha_{ox}$ 
out of this range of values (see  Figure~1 of \citealt{dors19}). 
 \item Metallicity: The values of metallicity in  relation to the solar one ($Z/Z_{\odot}$)= 0.2, 0.5, 0.75, 1.0, 1.5, and 2.0,
 were assumed in the models.  Assuming the solar oxygen abundance $\rm 12+\log(O/H)_{\odot}=8.69$ \citep{asplund09, allendeprieto01},
 the $Z$ values above corresponding to 12+log(O/H)= 8.0, 8.40, 8,56, 8.69, 8.86, 9.00, respectively,
Metallicity values in this range has been found for AGNs  with  
 redshifts  varying from $\sim 0$  to $\sim 7$ 
(e.g. \citealt{nagao06a, feltre16, matsuoka18, thomas19, mignoli19, enrique19, dors14, dors15, dors18}). 
 We found that photoionization models assuming ($Z/Z_{\odot}) \: > \: 2.0$ produce similar intensities of $N2$,  
 therefore, only  $(Z/Z_{\odot})\: \lid \: 2.0$ were assumed in our analysis.  The abundance of all  heavy elements was linearly
 scaled with $Z$, with the exception of the nitrogen abundance, which  was calculated by using the following relation
\begin{equation}
\label{non2}
\rm        
\log(N/O)=1.29\: \times \: 12+\log(O/H) \: - \: 11.84, 
\end{equation}
valid for $\rm 12+\log(O/H) \: \ga \: 8.0$ or $(Z/Z_{\odot}) \: \ga \: 0.2$.
This relation was obtained fitting  N and O abundance estimations derived using  
detailed  photoionization models by \citet{dors17} for a sample  of
Seyfert 2 AGNs located at $z \: < \: 0.1$ and  also taking   abundance estimations for \ion{H}{ii} regions into account.
The considered \ion{H}{ii} regions  are located  in irregular and 
spiral local galaxies and the oxygen abundance estimations were obtained  by \citet{leonid16}  using the C method \citep{leonid12}.
 In Figure~\ref{noohd17},  abundance estimations and the fit represented by the Equation~\ref{non2} are shown.

\begin{figure}
\centering
\includegraphics[angle=-90,width=1.0\columnwidth]{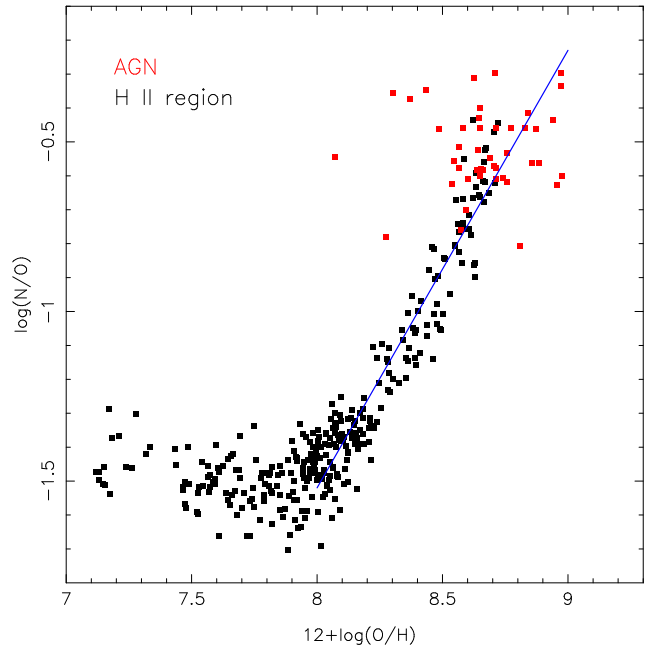}
\caption{log(N/O) versus 12+log(O/H) abundance ratio values. Red
points are values predicted by the individual photoionization models for  a 
sample of Seyfert~2 nuclei by \citet{dors17}. Black points are estimations
for \ion{H}{ii} regions   derived by   \citet{leonid16} through the   C method \citep{leonid12}.
The line represents a linear regression [for $\rm 12+\log(O/H) \: \ga \: 8.0$]
fitting to the points and represented by  Equation~\ref{non2}.}
\label{noohd17}
\end{figure}

It is worth to mention that the nitrogen and oxygen abundance relation  changes with the cosmic time (redshift) and any calibration
between $Z$ and   nitrogen emission-lines must  take into account the influence of this chemical evolution.
In fact, \citet{vincenzo18} analysed the evolution of the (N/O)-(O/H)  relation   with the redshift making use
of cosmological hydrodynamical simulations including detailed chemical enrichment. These authors
found  that higher N/O abundance ratios for a given O/H value are derived for low  redshift ($z \: \la \: 1$)  in comparison
with those having very high  redshift ($z \: \ga \: 5$, see Fig.~7 of their work). However, the  study 
carried out by \citet{vincenzo18} is based on star-forming regions modelling and, apparently, an opposite result is derived for 
AGN-like objects \citep{dors19}. 
Moreover, the (N/O)-(O/H)  relation  for star-forming regios can also change in cases where
these objects are located in   interacting galaxies \citep{koppen05, dors06}, 
although this has not been demonstrated  for AGNs.
Anyways, we emphasize that the $Z$-$N2$ relation derived  in this work
would be used  for studies of objects at  low redshift ($z \: \la \: 0.4$) and it must be applied with caution for objects at high redshift
and for AGNs in interacting galaxies.

 The internal presence of dust in the gas phase of gaseous nebulae has a strong influence on the emitted spectrum of these objects.
Dust grains absorb the ultraviolet radiation changing considerably the gas ionization degree. Moreover, dust grain  
 collision with gas atoms leads, in general, to a higher cooling rate of the gas, consequently, changing 
the emitted spectrum (e.g. \citealt{dwek92}). In particular,
the  effects of metal depletion  onto dust grains on the ionized gas of AGNs was analysed by
\citet{feltre16} finding that, when the dust-to-metal mass ratio increases, the removal of refractory coolant elements
from the gas phase reduces the cooling efficiency through infrared-fine structure transitions, 
implying in an increase of emission-lines emitted
by non-refractory elements, such as $N2$ (see also \citealt{kingdon95}). 
On the other hand, AGN models assuming the presence of dust in the gas phase  tend not to reproduce 
the majority of the ultraviolet emission-line intensities of AGNs \citep{nagao06a} and  even some authors have found difficulties 
in reproducing rest-frame optical or near-infrared emission lines of these objects (see \citealt{matsuoka09} and references
therein). Therefore, since the dust-to-metal mass ratio is poorly known in gaseous nebulae and AGNs \citep{peimbert10}
and, with the purpose of not introducing an additional uncertainty in our derived $Z$-$N2$ calibration, all the photoionization models considered in the present work are dust free.

\item Ionization parameter: The  ionization parameter ($U$) is defined as  $U= Q_{{\rm ion}}/4\pi R^{2}_{\rm in} N\, c$, where $ Q_{\rm ion}$  
is the number of hydrogen ionizing photons emitted per second by the ionizing source, $R_{\rm in}$  
is  the distance from the ionization source to the inner surface of the ionized gas cloud (in cm), $N$ is 
the  particle density (in $\rm cm^{-3}$), and $c$ is the speed of light (in $\rm km \: s^{-1}$).
We considered the logarithm of $U$  in the range of $ -4.0 \: \lid \: \log U \: \lid \: -0.5$, with a step of 0.5 dex, about 
the same values considered by  \citet{feltre16} for AGNs. A plane-parallel geometry was adopted and the outer radius was assumed to be the one where
the gas temperature reaches 4\,000 K (default outer radius value in the {\sc Cloudy} code).

\item Electron density: Three electron density values, constant along the NLR radius, were assumed in the models: 
$N_{\rm e}$= 100, 500 and 3000 $\rm cm^{-3}$. These values cover the $N_{\rm e}$ range derived for Seyfert~2 AGNs using the 
SDSS data \citep{vaona12, zhang13}.  
\end{enumerate}
In total, 399 photoionization models were built covering a wide range of AGN parameters.

\section{Results}
\label{res}

\begin{figure*}
\centering
\includegraphics[angle=-90,width=0.65\columnwidth]{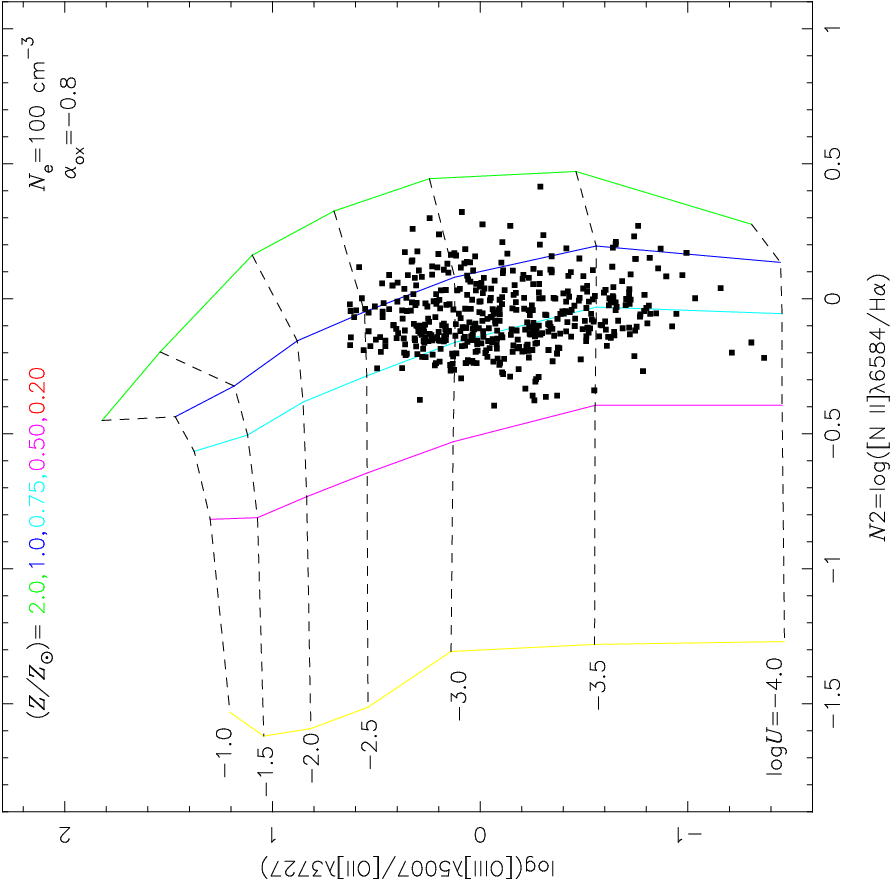}
\includegraphics[angle=-90,width=0.65\columnwidth]{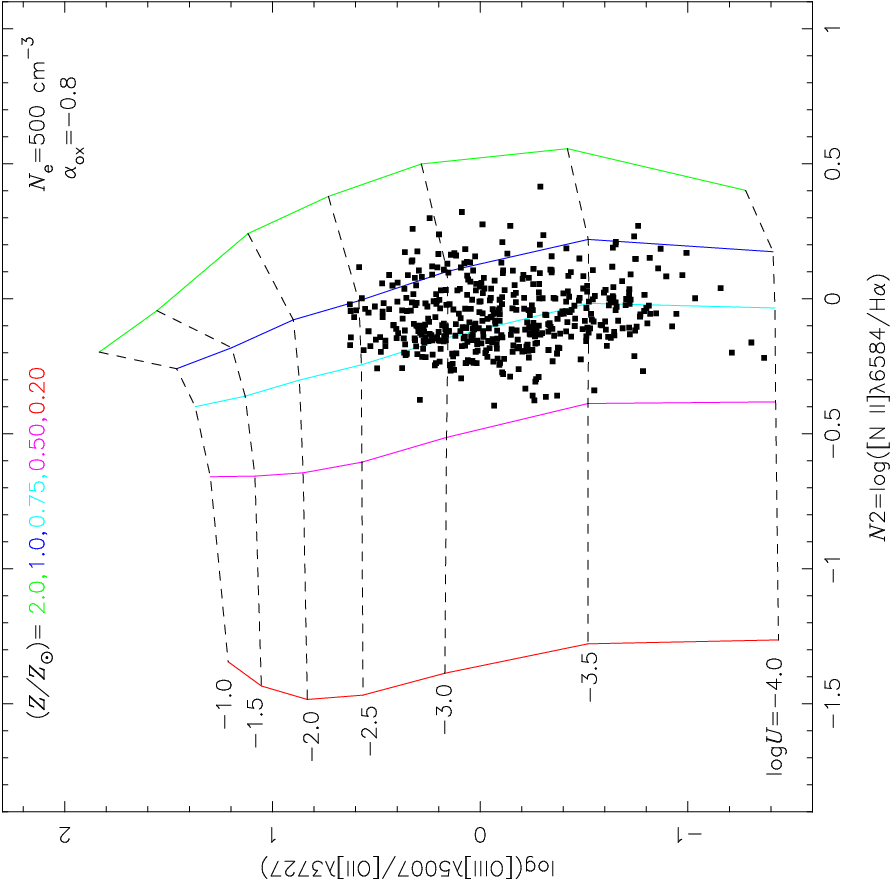}
\includegraphics[angle=-90,width=0.65\columnwidth]{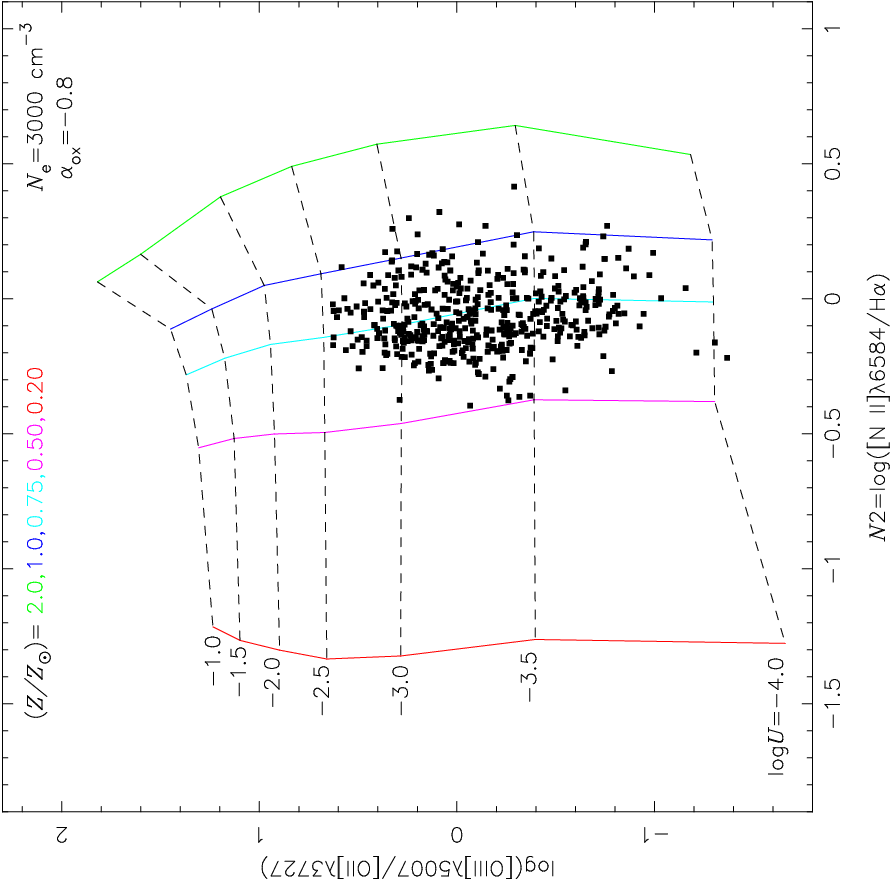}\\
\includegraphics[angle=-90,width=0.65\columnwidth]{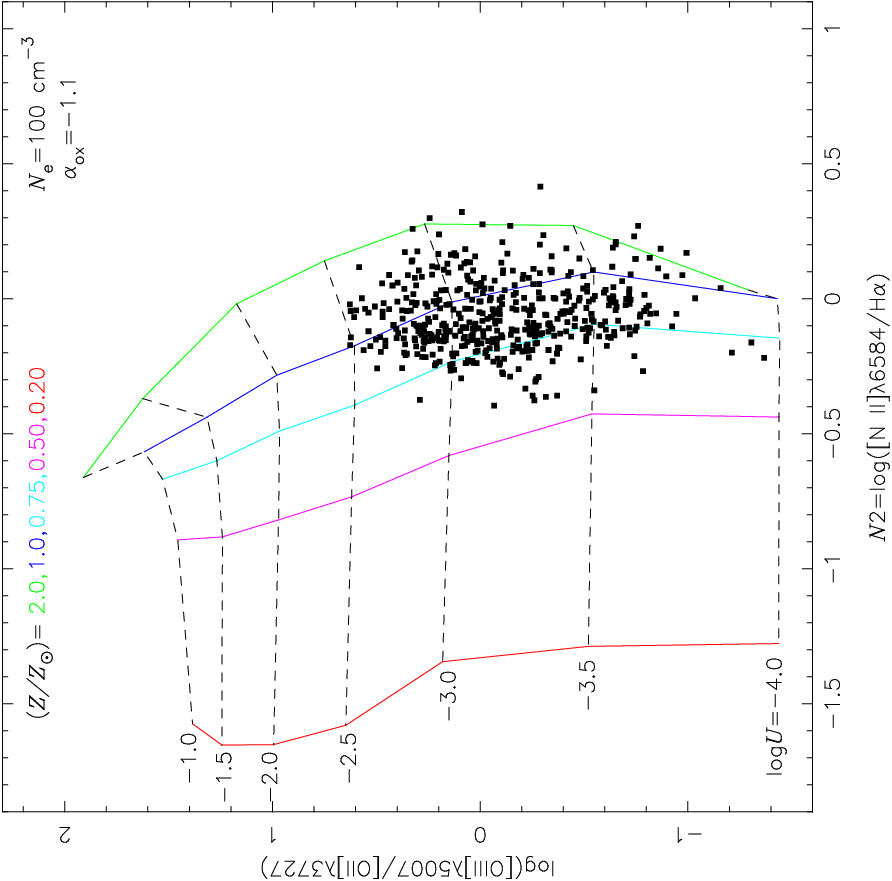}
\includegraphics[angle=-90,width=0.65\columnwidth]{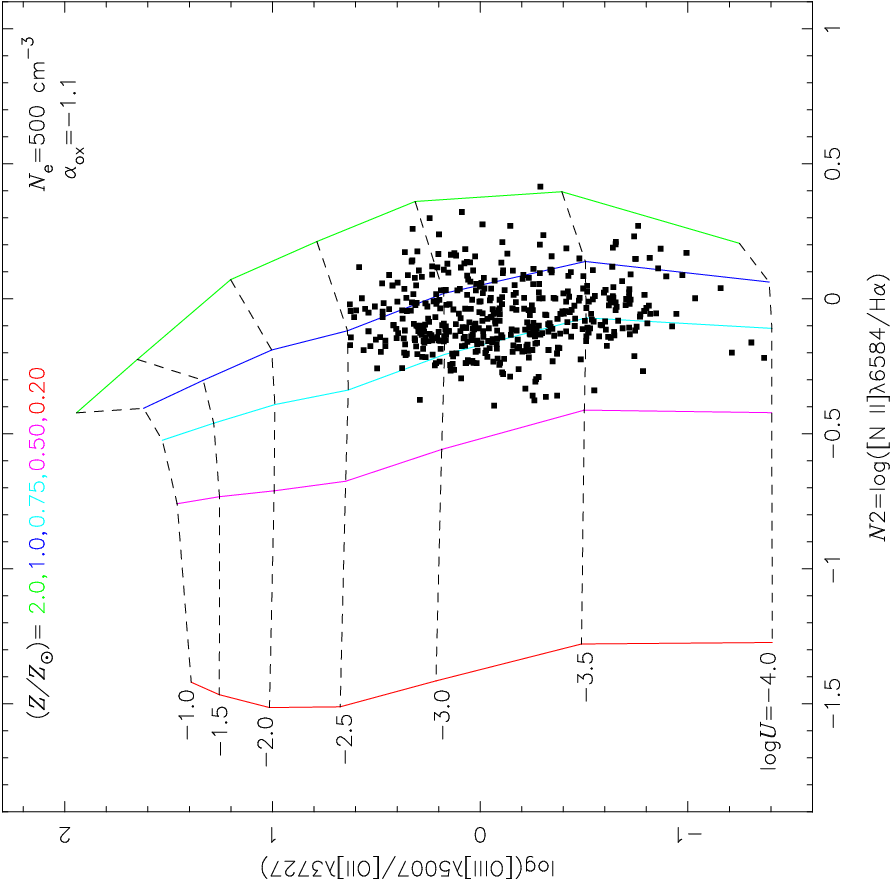}
\includegraphics[angle=-90,width=0.65\columnwidth]{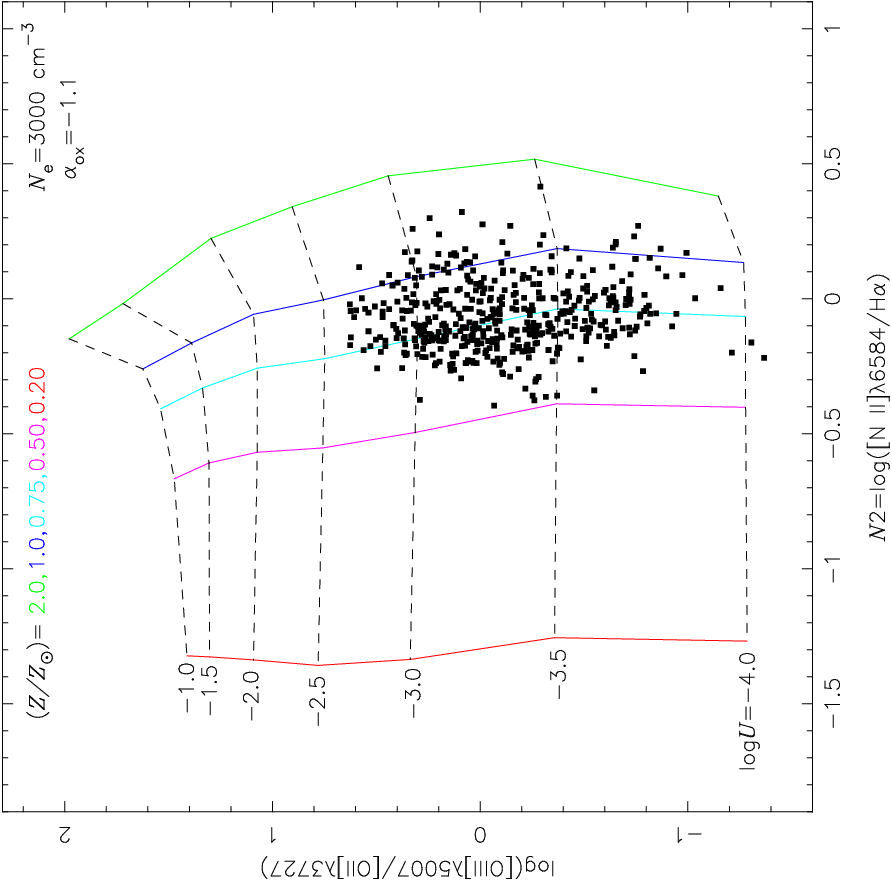}\\
\includegraphics[angle=-90,width=0.65\columnwidth]{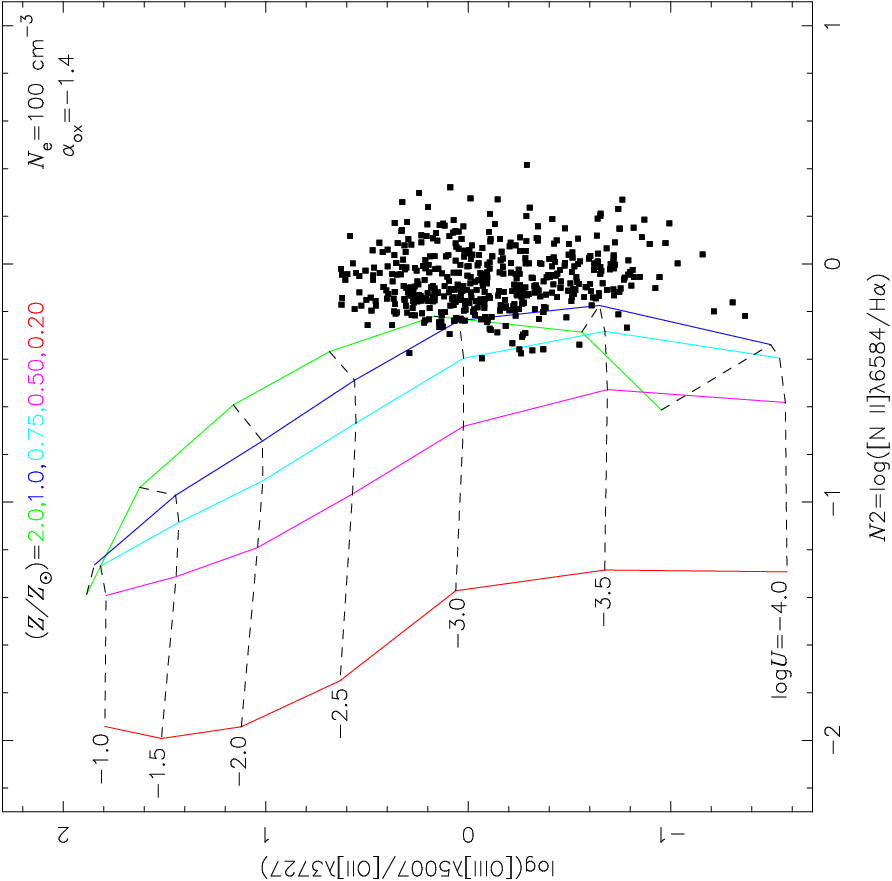}
\includegraphics[angle=-90,width=0.65\columnwidth]{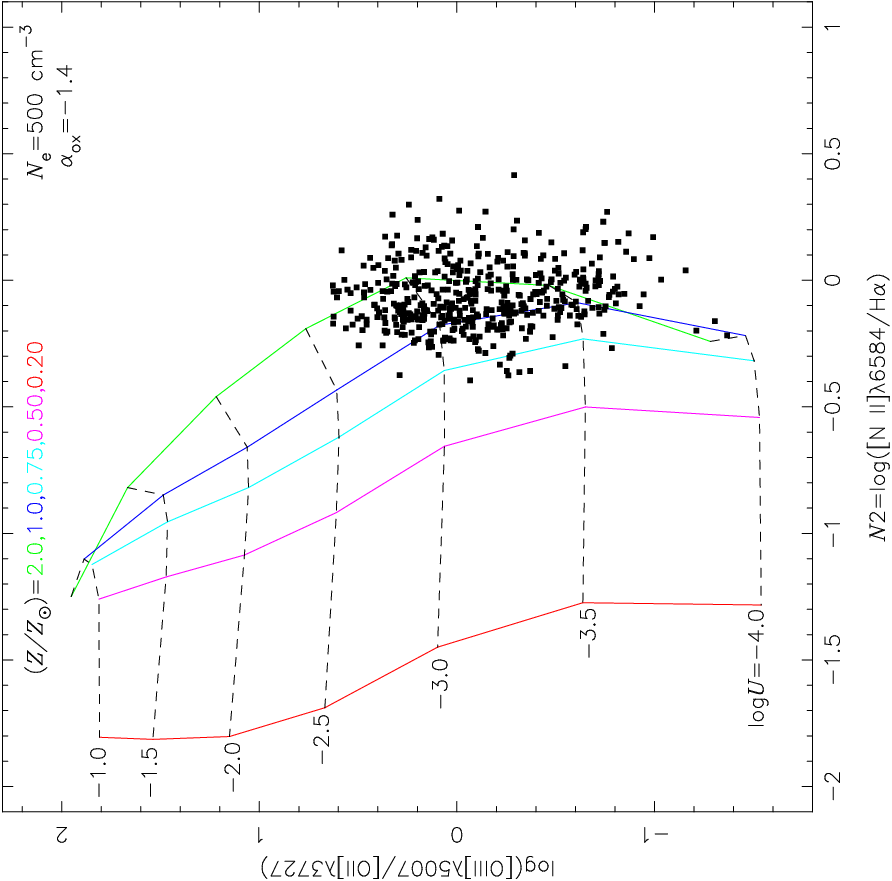}
\includegraphics[angle=-90,width=0.65\columnwidth]{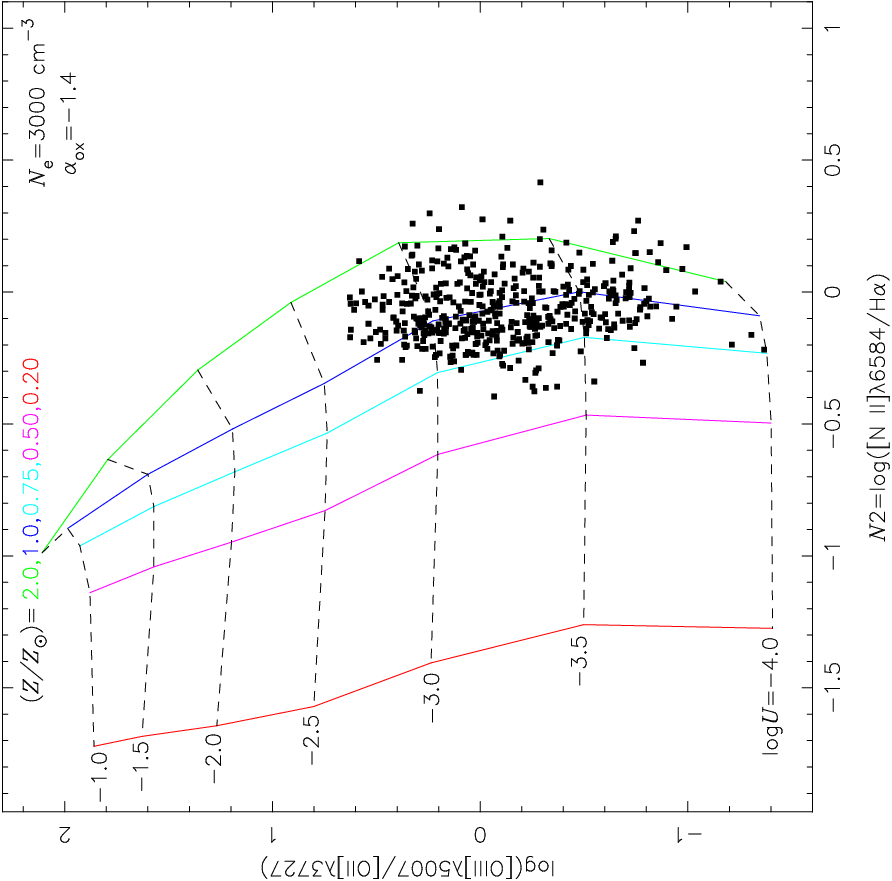}
\caption{log([\ion{O}{iii}]$\lambda5007$/[\ion{O}{ii}]$\lambda3727$) versus  $N2$=log([\ion{N}{ii}]$\lambda6584$/H$\alpha$).
Solid lines connect photoionization model results (see Sect.~\ref{mod}) with the same metallicity,  while
dashed lines connect models with same logarithm of the ionization parameter $U$, as indicated.
Points represent observational emission-line intensity ratios taken from the SDSS-DR7 (see Sect.~\ref{obs}).
In each plot, a grid of models assuming different electron density ($N_{\rm e}$)  and 
$\alpha_{ox}$ values, as indicated, are shown.}
\label{f2}
\end{figure*}

In Fig.~\ref{f2},  log([\ion{O}{iii}]$\lambda$5007/[\ion{O}{ii}]$\lambda$3727)
versus $N2$=log([\ion{N}{ii}]$\lambda$6584/H$\alpha$) diagrams containing the observational data and the photoionization model results
previously described are shown. Grids of models assuming distinct suppositions about $N_{\rm e}$ and $\alpha_{ox}$ values
are considered in each panel of Fig.~\ref{f2}. It is  plausible to note that photoionization models
with $N_{\rm e}$= 100, 500, 3000 $\rm cm^{-3}$ and $\alpha_{ox}=-0.8,-1.1$ well reproduce
 the observational data. As pointed out by \citet{groves04} and \citet{feltre16}, we found that
optical emission-line ratios are little sensitive to $N_{\rm e}$, under the collisional de-excitation
density  limit ($N_{\rm e} \: < 10^{4} \: \rm cm^{-3}$). 
However, when $\alpha_{ox}=-1.4$ is assumed, the models, in general, under-predict the
[\ion{N}{ii}]/H$\alpha$. 
Detailed photoionization modelling carried out by \citet{dors17} and bayesian-like comparison
between  Seyfert 2 optical emission lines and photoionization models by \citet{enrique19} also indicated that  
 $\alpha_{ox}\:  < \: -1.4$   are  representative of the SED of  Seyfert 2 AGNs.  
Therefore,  models with $\alpha_{ox}=-1.4$ are not considered in the derivation of the $Z$-$N2$ calibration.

\begin{figure}
\centering
\includegraphics[angle=-90,width=1.0\columnwidth]{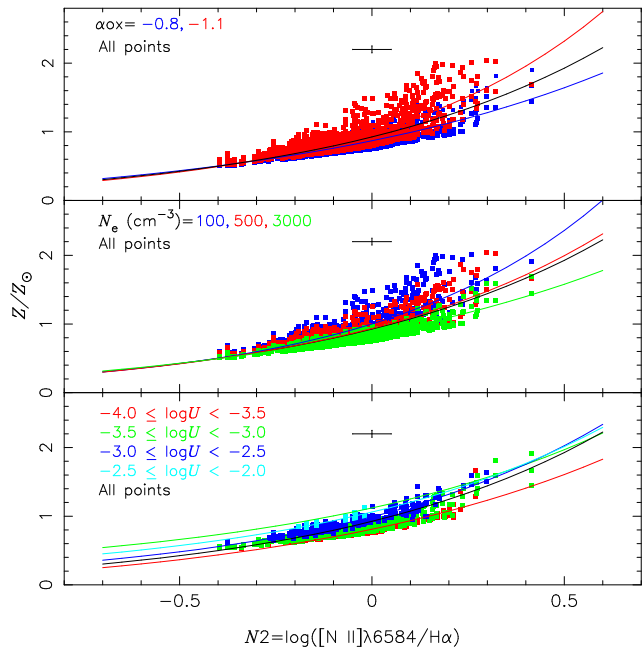}
\caption{Metallicity ($Z/Z_{\odot}$) versus the $N2$=log([\ion{N}{ii}]$\lambda$6584/H$\alpha$).
Points represent estimations obtained through linear interpolation between
 photoionization model results and the observational data 
presented in Fig.~\ref{f2}. Curves represent the fitting of the equation $(Z/Z_{\odot})=a^{N2}+b$ to the points
taking into account different model parameters (indicated in each panel),  whose the  coefficient fittings are
listed in Table~\ref{tab1}. Error bars in each panel represents the typical error (0.1 dex)  in  
observational measurements of the $N2$ index (e.g. \citealt{denicolo02}) and the 30\%
uncertainty in the interpolated values.}
\label{f3}
\end{figure}

To calibrate the metallicity as a function of the $N2$ index, 
we calculated the logarithm of the ionization parameter and the metallicity  
for each object of our sample by linear interpolations between  the
models shown in Fig.~\ref{f2}. The typical error in emission-line ratio intensities is about 0.1 dex 
(e.g. \citealt{denicolo02, kenniccutt03}). Assuming
this uncertainty in the data considered in Fig.~\ref{f2}, we obtained an uncertainty in the $Z$ and $\log U$ 
interpolated estimations in order of 30\% and 0.05 dex, respectively.
In the panels of Fig.~\ref{f3},  the relation between $Z/Z_{\odot}$ and $N2$, 
considering models with distinct  $N_{\rm e}$ and $\alpha_{ox}$ values and ranges of $\log U$ are shown. 
We use the following expression: 
\begin{equation}
\label{eq1}
(Z/Z_{\odot})=a^{N2}+b
\end{equation} 
to fit the results obtained for the objects in our sample  plotted
in  Fig.~\ref{f3}. The fitting coefficients are listed in Table~\ref{tab1}.
As it can be seen, the correlation of the derived parameters with $\log U$
is marginal, indicating a very low dependence of the $Z-N2$ relation on
the ionization degree in the AGN.
On the other hand, a larger dependence of the $Z$-$N2$ relation 
on $N_{\rm e}$ is found, in the sense that higher (up to a
factor of 2) $Z$ estimations are obtained  when photoionization models with  lower $N_{\rm e}$ are considered, mainly
for the high metallicity regime $[(Z/Z_{\odot}) \: \ga \: 1.0]$. 
 Similarly,   a dependence of $Z$-$N2$ on $\alpha_{ox}$
is also derived, in the sense that higher metallicity (up to a
factor of 2) is derived if $\alpha_{ox}=-1.1$ is assumed in comparison with those considering $\alpha_{ox}=-0.8$,
being the difference between the estimations also more  prominent for $(Z/Z_{\odot}) \: \ga \: 1.0$.
Interestingly, an opposite behaviour was found by \citet{dors19} for the relation between $Z$ and ultraviolet emission-line ratios
(see also \citealt{nagao06a}). 
We also fitted  Eq.~\ref{eq1} considering all points (not discriminating  nebular parameters) and the resulting coefficients are listed
in Table~\ref{tab1}.

\begin{table}
\caption{Values of the $a$ and $b$ coefficients resulting from  fittings of the Eq.~\ref{eq1}  to the
estimations, shown in Fig.~\ref{f3}, for different model parameters.  The last line lists
the coefficients obtained not discriminating the model parameters.}
\label{tab1}
\begin{tabular}{ccc}
\hline
 Model parameter       	      &	      $a$	 &	 $b$	     \\
\hline
$\log U$                      &     		 &                \\
($-$4.0, $-$3.5) 	      & $3.23\pm0.11$    & $-0.19\pm0.01$ \\	  	    
($-$3.5, $-$3.0)	      & $3.42\pm0.06$    & $-0.12\pm0.01$ \\	  		    
($-$3.0, $-$2.5)	      & $4.15\pm0.12$    & $-0.01\pm0.01$ \\	  		    
($-$2.5, $-$2.0)	      & $3.82\pm1.01$    & $+0.06\pm0.02$ \\	  	    
 \hline
 $N_{\rm e}$($\rm cm^{-3}$)   &                  & 	          \\
 100                          &$5.58\pm0.23$	 & $+0.00\pm0.01$ \\ 
 500			      &$4.24\pm0.12$	 & $-0.06\pm0.01$ \\ 
 3000			      &$2.99\pm0.04$	 & $-0.15\pm0.01$ \\
\hline
 $\alpha_{ox}$                &                  &	            \\
 $-0.8$                       & $3.14\pm0.05$	 & $-0.13\pm0.01$   \\ 
 $-1.1$                       & $5.45\pm0.17$	 & $-0.01\pm0.01$   \\
 \hline				
All points                    & $4.01\pm0.08$    & $-0.07\pm0.01$   \\
\hline
\end{tabular}
\end{table}

\begin{figure}
\centering
\includegraphics[angle=-90,width=1.0\columnwidth]{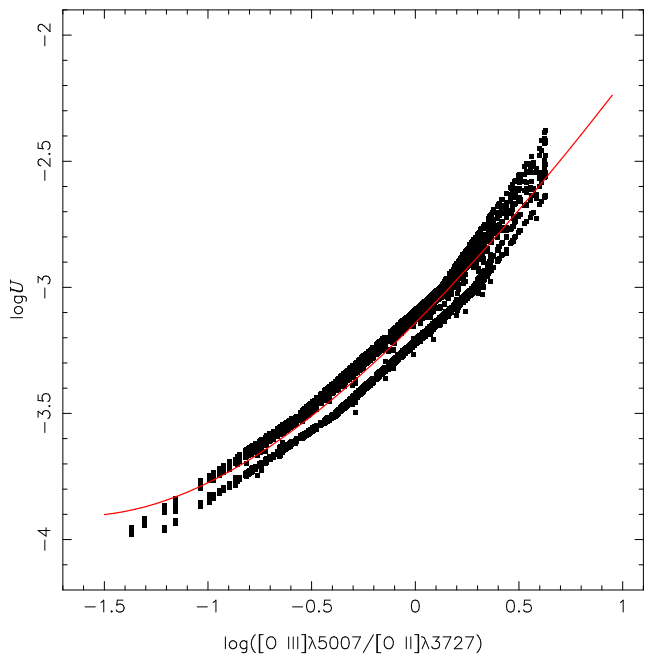}
\caption{As in Fig.~\ref{f3} but for the logarithm of the ionization parameter ($\log U$) versus the  logarithm
of the line ratio [\ion{O}{iii}]$\lambda$5007/[\ion{O}{ii}]$\lambda$3727. Photoionization model results assuming
different parameters are not discriminated. Curve represents the fitting to the points and represented by  Eq.~\ref{eq2}.
The error in the interpolated $\log U$ values is about 0.05 dex.}
\label{f4}
\end{figure}

The interpolated values from Fig.~\ref{f2} made it possible  to derive a relation 
between  the logarithm of the ionization parameter and [\ion{O}{iii}]/[\ion{O}{ii}] line ratio, 
shown in Fig.~\ref{f4}. The linear regression obtained is:
\begin{equation}
\label{eq2}
\log U=(0.57\pm 0.01 \: x^{2})+(1.38\pm 0.01 \: x) - (3.14\pm 0.01),
\end{equation}
where $x$=log([\ion{O}{iii}]$\lambda$5007/[\ion{O}{ii}]$\lambda$3727).
We did not find any dependence of this  equation on $N_{\rm e}$,
$\alpha_{ox}$ and $Z$.

\section{Discussion}
\label{disc}  

\citet{thaisa94} proposed the use of the $N2$=log([\ion{N}{ii}]$\lambda$6584/H$\alpha$) line ratio as  an indicator
of the  ratio between oxygen and hydrogen abundances of \ion{H}{ii} regions. These authors obtained a calibration 
based on  O/H abundances calculated through the $T_{\rm e}$-method and  observational emission-line intensities of  star-forming
galaxies. Thereafter, other authors \citep{raimann00, denicolo02, pettini04, liang06, grazina06, nagao06b, yin07, enrique09, 
marino13, moralesluis14}  improved this calibration  by including  more abundance estimations, mainly for 
both  low and  high metallicity ends. The advantage of the $N2$ index  over the commonly used  metallicity indicator $R_{23}$ is that:
($i$) it does not include the [\ion{O}{ii}]$\lambda$3727 line, which makes this line ratio not sensitive to reddening correction and,
consequently, useful to dusty object studies (e.g. \citealt{xiao12}); ($ii$) due to the fact that $N2$ involves emission-lines 
with very close  wavelength, it is not affected by uncertainties of flux calibration
\citep{marino13}, ($iii$) it is accessible in the near infrared at moderate-to-high redshifts (e.g. \citealt{cresci12, queyrel12}),
 ($iv$) it has a   less critical dependence on the ionization parameter,  ($v$) it is single-valued with $Z$, and
 ($vi$) it has a tighter correlation with O/H \citep{denicolo02}. 

Despite the several advantages, such as other $Z$ indicators, $N2$ index suffers some limitations.
Firstly, for any theoretical  calibration involving nitrogen lines it is necessary to know the 
dependence between  N/O and O/H abundance ratios (see \citealt{enrique09}). 
For AGNs, this relation was first derived 
by \citet{dors17}, who used detailed photoionization model  of  relatively  small (44 objects) 
sample of local ($z \: < \: 0.1$) Seyfert 2 AGNs (see also \citealt{enrique19}).
Obviously, it is necessary to  obtain  N and O abundance estimations for a larger sample of objects at  a wider redshift range.
Moreover, the dependence between the  nitrogen lines and $Z$ (or O/H) is due to the N secondary stellar 
nucleosynthesis origin [${\rm (N/O)} \approx Z^{2}$] in the ``high" metallicity 
regime [$(Z/Z_{\odot})\: \ga \: 0.3$] (e.g. \citealt{vilacostas93}). Therefore, calibrations
involving nitrogen lines are not valid for the low metallicity regime.
Finally, the $N2$ index saturates in the very high-metallicity regime \citep{marino13}, as it is reported in Sect.~\ref{res}.
In the case of our   $Z$-$N2$ calibration, it is valid for the range of $0.3 \: \la \: (Z/Z_{\odot}) \: \la \: 2.0$,
 which  corresponds to $-0.7 \: \la \: (N2) \: \la \: 0.6$.

Regarding  the $Z$-$N2$ calibration dependence on the electron density ($N_{\rm e}$), we found that  it is 
more prominent  in the high metallicity regime [$(Z/Z_{\odot}) \: \ga 1.0$]. 
Although $N_{\rm e}$ is easily estimated in AGNs through 
the dependence of this parameter with the [\ion{S}{ii}]$\lambda$6716$\lambda$/6731 line ratio, the
observational measurement  error of $N2$ ($\sim 0.1$ dex , \citealt{denicolo02}) translates in a $Z$ 
uncertainty of the order of the one obtained not taking into account the  $N_{\rm e}$ effects on our calibration.
It can be seen in Fig.~\ref{f3}, where the typical error of $N2$ is shown in the panels. 
The same result is derived for the effect of $\alpha_{ox}$ on the metallicity estimations, which the
$Z$ uncertainty of not knowing $\alpha_{ox}$ is of the order of the  uncertainty produced by the observational $N2$ error.  
It is worth to mention that similar results were derived by \citet{dors19}. These authors showed that the uncertainties in  $Z$ estimations assuming
photoionization models with different $N_{\rm e}$ and $\alpha_{ox}$ values are similar to  those
produced by observational errors of ultraviolet emission-line ratios (see Fig.~5 of their work).

Recently,   in  Paper~I, we compared the AGN Seyfert~2 metallicities (traced by the O/H abundance ratio) and the mass-metallicity relation
derived by using most of the available methods in the literature and this analysis
will not be repeated here. For simplicity and with the goal to validate our $Z$-$N2$ calibration,
we only compare estimations for our sample by using  Eq.~\ref{eq1} 
with those derived by using two calibrations involving nitrogen emission lines, i.e. 
the first calibration of \citet{thaisa98} and the \citet{castro17} calibration as well
as  results from  bayesian-like approximation proposed by \citet{enrique19}.

The $N2$ index combined with [\ion{O}{iii}]/H$\beta$ and [\ion{O}{iii}]/[\ion{O}{ii}] line ratios 
 was proposed as  O/H abundance indicator of the  NLR of AGNs by \citet{thaisa98}. These
authors proposed two theoretical calibrations based on a grid of photoionization models assuming 
the (N/O)-(O/H) relation derived for nuclear starbursts by \citet{thaisa94} and given by the relation
\begin{equation}
\label{nothaisa}
\rm \log(N/O)=[0.96\: \times \: (12+\log(O/H)]-9.29.
\end{equation}  
 In  Paper~I, we found that both calibrations of \citet{thaisa98}  produce very similar results (with an average difference 
of $-0.08$ dex). Therefore, we will consider only the first calibration of these authors given by
\begin{eqnarray}
       \begin{array}{l@{}l@{}l}
\rm (O/H)_{SB98,1} & = &  8.34  + (0.212 \, x) - (0.012 \,  x^{2}) - (0.002 \,  y)  \\  
         & + & (0.007 \, xy) - (0.002  \, x^{2}y) +(6.52 \times 10^{-4} \, y^{2}) \\  
         & + & (2.27 \times 10^{-4} \, xy^{2}) + (8.87 \times 10^{-5} \, x^{2}y^{2}),   \\
     \end{array}
\label{sb1}
\end{eqnarray}
\noindent where $x$ = [N\,{\sc ii}]$\lambda$$\lambda$6548,6584/H$\alpha$ and 
$y$ = [O\,{\sc iii}]$\lambda$$\lambda$4959,5007/H$\beta$.
 The term O/H above corresponds to 12+log(O/H)  and it is converted into metallicity by 
\begin{equation} 
(Z/Z_{\odot})=10^{8.69-\rm (O/H)_{SB98,1}},
\end{equation}
being 8.69 dex the solar oxygen abundance \citep{asplund09, allendeprieto01}.
The calibration above is valid for   $\rm 8.4 \: \lid \: 12+log(O/H) \:  \lid \: 9.4$.
A correction in the O/H derivation  due to the electron density effects  on the calibration
above is given by 
\begin{equation}
{\rm (O/H)_{final}=[(O/H)}-0.1 \: \times \: \log (N_{\rm e}/300 ({\rm cm^{-1}}))].
\end{equation}
Another calibration for AGNs  involving [\ion{N}{ii}] lines was proposed by \citet{castro17}  
  considering $N2O2$ index. These authors assumed in the photoionization models
 the following (N/O)-(O/H) relation  derived for  star-forming regions by \citet{dopita00}:
  \begin{eqnarray}
       \begin{array}{lll}
 \log({\rm N/H})   & = &  -4.57+\log(Z/Z_{\odot}); \: {\rm for} \:  \log(Z/Z_{\odot}) \lid -0.63,   \\  
    \log({\rm N/H}) & = & -3.94+2\: \log(Z/Z_{\odot});  \: {\rm otherwise}. \\  
     \end{array}
\label{non2o2}
\end{eqnarray}
The calibration derived by \citet{castro17} is
 \begin{eqnarray}
     \begin{array}{lll}
(Z/Z_{\odot}) \!\!\!  & =  &\!\!\! 1.08(\pm 0.19) \times N2O2^2  +  1.78(\pm0.07) \times N2O2  \\
                      &    &           +1.24(\pm0.01) . \\  
     \end{array}
\label{zcastro}
\end{eqnarray}

The  bayesian-like {\sc \ion{H}{ii}-Chi-mistry} code  (hereafter {\sc HCm}, \citealt{enrique14})
was used to estimate the O/H and N/O abundance ratios   of each object
of the sample described in Sect.~\ref{obs}. The  {\sc HCm} code is based on a bayesian-like 
comparison between certain observed emission-line ratios sensitive to total oxygen abundance, 
nitrogen-to-oxygen ratio, and ionization parameter with the predictions from a large grid of 
photoionization models. The  {\sc HCm} code does not consider a fixed (N/O)-(O/H) relation. 
In \citet{enrique19} this code was adapted for AGNs. 
 
\begin{figure}
\centering
\includegraphics[angle=-90,width=1.0\columnwidth]{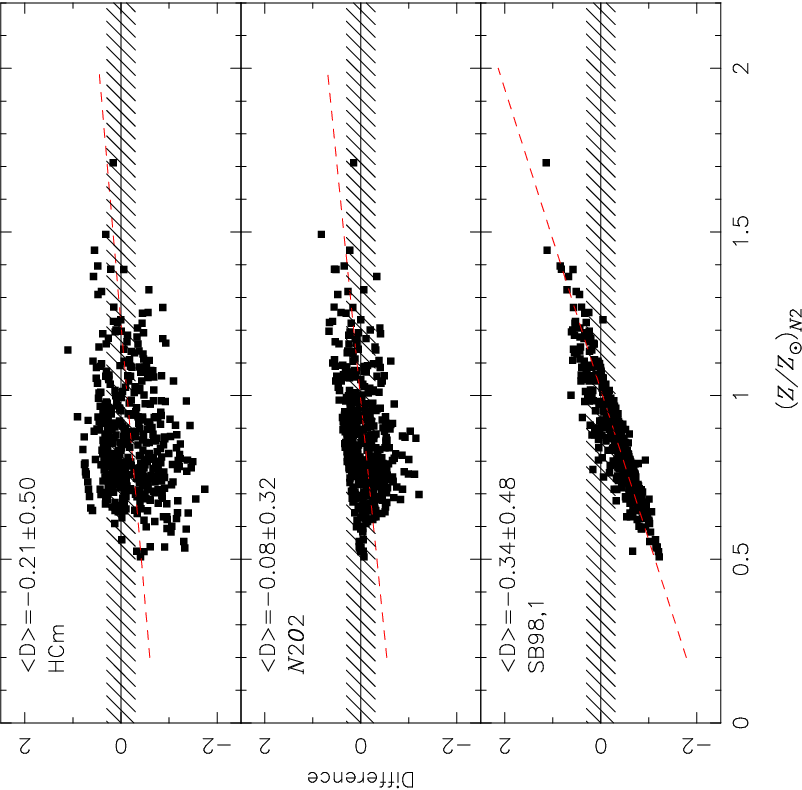} 
\caption{Bottom panel: Difference between the metallicity estimations for our sample of
objects (see Sect.~\ref{obs}) obtained from our $Z$-$N2$ calibration (Eq.~\ref{eq1})
and those from \citet{thaisa98}  calibration  
versus the  $Z$-$N2$ estimations. Middle panel: As the bottom panel but for $Z$-$N2O2$ calibration
(Eq.~\ref{zcastro}) proposed by \citet{castro17}. Upper panel: As the bottom panel but for
estimations obtained by using the {\sc HCm} code \citep{enrique19}. In each panel,
the average and standart desviation of the difference between  the estimations are shown. The dashed area indicates
the uncertainty of $\pm 0.1$ dex assumed in  $Z$ estimations via strong emission-line methods \citep{denicolo02}.}
\label{f5}
\end{figure}

 In Fig.~\ref{f5}, the differences between the estimations via our $N2$ calibration (Eq.~\ref{eq1}) and those
 via the calibrations proposed by \citet{thaisa98} and \citet{castro17}  as well as  those derived using the {\sc HCm} code
 are plotted against the  estimations via Eq.~\ref{eq1}. The estimations via our calibration (Eq.~\ref{eq1}) were obtained assuming the fitting for 
all model results, i.e. all $Z-N2$ values, whose coefficients are listed in Table~\ref{tab1}.
  It can be seen in Fig.~\ref{f5} that a systematic difference
is found between the estimations based on our $N2$ calibration and those via \citet{thaisa98} calibration, in the sense that 
the latter calibration produces lower and higher $Z$ values for the low and high metallicity regime, respectively.
Although  similar results have been derived for the difference between the estimations by using the
$N2O2$ calibration and those via  HCm code,  these are less prominent  than the one obtained by using \citet{thaisa98} calibration.

\begin{figure}
\centering
\includegraphics[angle=-90,width=1.0\columnwidth]{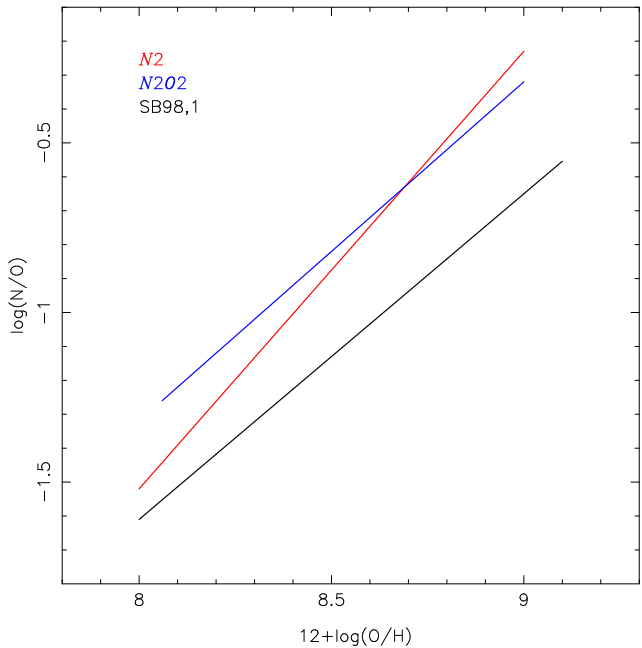} 
\caption{Comparison between the (N/O)-(O/H) relations (represented by the lines) assumed in the photoionization models
 in this paper (Eq.~\ref{non2}), by \citet{thaisa98} (Eq.~\ref{nothaisa}), and by \citet{castro17} (Eq.~\ref{non2o2})  
 to obtain  the calibrations  represented by the Eqs.~\ref{eq1},  \ref{sb1} and \ref{zcastro}, respectively.}
\label{f6}
\end{figure}

The differences in the $Z$ estimations found in Fig.~\ref{f5} are probably due to 
the use of distinct (N/O)-(O/H) relation in the  photoionization models used to obtain the calibrations. 
In order to verify that, in Fig.~\ref{f6}, the (N/O)-(O/H) relations used in the photoionization models 
to obtain the calibrations considered in Fig.~\ref{f5} are shown.  It can be seen that
the (N/O)-(O/H) relation assumed by us in this paper (Eq.~\ref{non2}) and by \citet{castro17} (Eq.~\ref{non2o2}) are very similar to each other, 
 clarifying the lowest $Z$  difference found in Fig.~\ref{f5}. On the other hand, 
the relation used by \citet{thaisa98} (Eq.~\ref{nothaisa}) produces lower N/O abundances  in comparison with those 
from the relations assumed in the $N2$ and $N2O2$ calibrations.
   
\begin{figure}
\centering
\includegraphics[angle=-90,width=1.0\columnwidth]{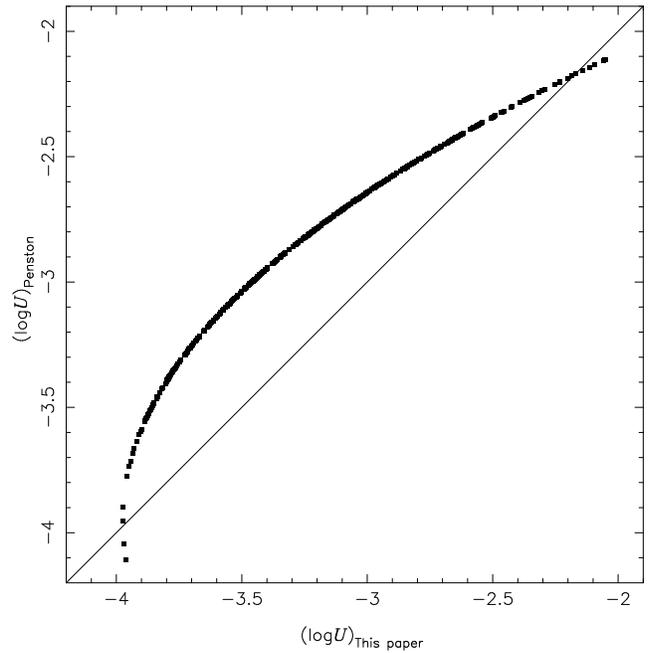} 
\caption{Logarithm of the ionization parameter ($\log U$) 
for our sample (see Sect.~\ref{obs}) derived
by  using the Eq.~\ref{ionpen} proposed by \citet{penston90} versus those calculated from our calibration
(Eq.~\ref{eq2}). The line represents the equality between the estimations.}
\label{f7}
\end{figure}

Regarding  the ionization parameter, few authors have proposed a calibration between $U$ and
narrow optical line-ratios of AGNs. For instance, \citet{penston90} proposed a calibration
between $U$ and the [\ion{O}{ii}]$\lambda$3727/[\ion{O}{iii}]$\lambda$5007 line ratio. These authors
used  sequences of photoionization models, taken from \citet{robinson87}, employing a variety of possible 
SEDs for the ionizing source and
assuming only one value of electron density ($N_{\rm e}=100 \: \rm cm^{-3}$) and solar metallicity.
The relation derived by  \citet{penston90} is
\begin{equation}
\label{ionpen}
\log U =-2.74 - y,
\end{equation}
where $y$=log([\ion{O}{ii}]$\lambda$3727/[\ion{O}{ii}]$\lambda$5007.
Hence the $U$ definition assumed in \citet{robinson87} is equal to the 
one of our models, it is possible  to compare estimations derived from their calibration  with
the ones obtained from our calibration. 
In Fig.~\ref{f7}, the logarithm of  the ionization parameter ($\log U$) calculated
by using the Eq.~\ref{ionpen} for our sample of objects are compared to 
those via our calibration (Eq.~\ref{eq2}).  It can be seen that, in general, 
the \citet{penston90} calibration produces somewhat higher $\log U$ values
than those derived from our calibration. This discrepancy, probably,  is due to
the calibration proposed by \citet{penston90} was obtained by using  photoionization models with fixed values
of $N_{\rm e}$ and $Z$, while in our calibration  a semi-empirical aproximation is considered, taken into
account a large range of nebular parameter.

\section{Summary and conclusions}
\label{conc}
We combined results of photoionization model built with the {\sc Cloudy} code with
observational data of 463 confirmed Seyfert 2 nuclei (redshift $z \: \la 0.4$), taken from the Sloan Digital Sky Survey DR7 dataset,
in order to obtain a semi-empirical calibration between the metallicity ($Z$) of the Narrow Line Region
of these objects and  the  $N2$=log([\ion{N}{ii}]$\lambda$6584/H$\alpha$) emission-line intensity ratio.
Our $Z$-$N2$ relation  is valid for the range  of $0.3 \: \la \: (Z/Z_{\odot}) \: \la \: 2.0$,  which
corresponds to $-0.7 \: \la \: (N2) \: \la \: 0.6$. The effects of varying
the  ionization parameter ($U$), electron density and the slope of the Spectral Energy Distribution
on the $Z$ estimations are  of the order of the uncertainty produced by the error measurements of $N2$.  
This result indicates the large reliability of our $Z-N2$ calibration. We also derived a calibration
between $\log U$ and the line ratio [\ion{O}{iii}]$\lambda$5007/[\ion{O}{ii}]$\lambda$3727,  
less dependent on other nebular parameter.

\section*{Acknowledgments}
This work is partially supported by the Brazilian agencies FAPESP, CAPES and CNPq.
EPM acknowledges support from the Spanish MINECO
project  Estallidos 6 AYA2016-79724-C4.
and  by the Spanish Science Ministry "Centro de Excelencia Severo Ochoa Program
under grant SEV-2017-0709.

% =============================================================================================================

\label{lastpage}


\begin{thebibliography}{99}
\bibitem[Abazajian et al.(2009)]{abazajian09} Abazajian K.~N., Adelman-McCarthy J.~K., Ag\"ueros M.~A. et al., 2009, ApJS, 182, 543 
\bibitem[Alende Prieto et al.(2001)]{allendeprieto01} Alende Prieto C., Lambert D.~L., Asplund M., 2001, ApJ, 556, L63
\bibitem[Asplund et al.(2009)]{asplund09} Asplund M., Grevesse N., Sauval A.~J., Scott P., 2009, ARA\&A, 47, 481
\bibitem[Baldwin et al.(1981)]{baldwin81} Baldwin J.~A., Phillips M.~M., Terlevich R., 1981, PASP, 93, 5
\bibitem[Castellanos et al.(2002)]{castellanos02} Castellanos M., D\'{\i}az A.~I., Terlevich E., 2002, MNRAS, 337, 540 
\bibitem[Castro et al.(2017)]{castro17} Castro C. S., Dors O.~L., Cardaci M.~V., H{\"a}gele G.~F., MNRAS, 467, 1507
\bibitem[Contini(2017)]{contini17} Contini M., 2017, MNRAS, 469, 3125
\bibitem[Contini \&  Aldrovandi(1983)]{contini83} Contini M., \&  Aldrovandi S.~M.~V., 1983, A\&A, 127, 15 
\bibitem[Cresci et al.(2012)]{cresci12} Cresci G., Mannucci F., Sommariva V., et al. 2012, MNRAS, 421, 262
\bibitem[Denicol\'o et al.(2002)]{denicolo02} Denicol\'o G., Terlevich R., Terlevich E., 2002, MNRAS, 330, 69
\bibitem[Dwek \& Arendt(1992)]{dwek92} Dwek E., Arendt R.~G., 1992, ARA\&A, 30, 11
\bibitem[Dopita et al.(2000)]{dopita00} Dopita M.~A., Kewley L.~J., Heisler C.~A., Sutherland R.~S., 2000, ApJ, 542, 224
\bibitem[Dors et al.(2012)]{dors12} Dors O.~L., Riffel R.~A., Cardaci M.~V. et al., 2012, 422, 252
\bibitem[Dors et al.(2014)]{dors14} Dors O.~L., Cardaci M.~V., H\"agele G.~F.,  Krabbe \^A.~C., 2014, MNRAS, 443, 1291
\bibitem[Dors et al.(2015)]{dors15} Dors O.~L., Cardaci M.~V., H\"agele G.~F., Rodrigues I., Grebel E.~K., Pilyugin, L.~S., Freitas-Lemes, P., Krabbe \^A.~C., 2015, MNRAS, 453, 4102
\bibitem[Dors et al.(2017)]{dors17} Dors O.~L., Arellano-C\'ordoba K.~Z., Cardaci M.~V., H\"agele G.~F., 2017, 468, L113
\bibitem[Dors et al.(2018)]{dors18} Dors O.~L., Agarwal B., Hägele G.~F., Cardaci M.~V., Rydberg C., Riffel R.~A., Oliveira A.~S., Krabbe A.~C., 2018, MNRAS, 479, 2294
\bibitem[Dors et al.(2019)]{dors19} Dors O.~L., Monteiro A.~F., Cardaci M.~V., H\"agele G.~F., Krabbe \^A.~C., 2019, MNRAS, 489, 241
\bibitem[Dors et al.(2020)]{dors20} Dors O.~L., Freitas-Lemes P, \^Amores E.~B. et al., 2020, MNRAS, 492, 468,  Paper~I 
\bibitem[Dors \& Copetti(2006)]{dors06} Dors O.~L., \& Copetti M.~V.~F, 2006, A\&A, A\&A 452, 473 
%\bibitem[Kewley et al.(2001)]{kewley01} Kewley L.~J., Dopita M.~A., Sutherland R.~S., Heisler C.~A., Trevena J., 2001, ApJ, 556, 121
\bibitem[Kennicutt et al.(2003)]{kenniccutt03} Kennicutt R.~C., Bresolin F., Garnett D. R., 2003, ApJ, 591, 801
\bibitem[Kingdon, Ferland \& Feibelman(1995)]{kingdon95} Kingdon J., Ferland G.~J., Feibelman W.~A., 1995, ApJ, 439, 793
\bibitem[H\"agele et al.(2008)]{hagele08} H\"agele, G. F., D\'az, A.~I., Terlevich, E., Terlevich, R., P\'erez-Montero, E., Cardaci, M.~V. 2008, MNRAS, 383, 209 
\bibitem[Ho(1999)]{ho99} Ho L.~C., 1999, ApJ, 516, 672
\bibitem[Izotov et al.(2006)]{izotov06} Izotov Y. I., Stasi\'nska G., Meynet G., Guseva N. G., Thuan T. X., 2006, A\&A, 448, 955
\bibitem[Feltre, Charlot \& Gutkin(2016)]{feltre16} Feltre  A., Charlot S., Gutkin J., 2016, MNRAS, 456, 3354
\bibitem[Ferland et al.(2017)]{ferland17} Ferland G.~J. et al., 2017, Rev. Mex. Astron. Astrofis., 53, 385
\bibitem[Garcia-Rojas(2020)]{jorge20} Garc{\'\i}a-Rojas J., 2020, arXiv:2001.03388
\bibitem[Groves et al.(2004)]{groves04} Groves B.~A., Dopita M.~A., Sutherland R., 2004, ApJ, 153, 75
\bibitem[Groves et al.(2006)]{groves06} Groves B.~A., Heckman T.~M., Kauffmann G., 2006, MNRAS, 371, 1559
\bibitem[Gusev et al.(2012)]{gusev12} Gusev A.~S., Pilyugin L.~S., Sakhibov F., Dodonov S.~N., Ezhkova O.~V., Khramtsova M.~S., 2012, MNRAS, 424, 1930
\bibitem[do Nascimento et al.(2019)]{janaina19} do Nascimento J.~C., Storchi-Bergmann T., Mallmann N.~D. et al., 2019, 486, 5075
\bibitem[Kennicutt et al.(2003)]{kennicutt03} Kennicutt R. C., Bresolin F., Garnett D. R., 2003, ApJ, 591, 801
\bibitem[Kewley et al.(2019)]{kewley19} Kewley  L.~J., Nicholls D.~C., Sutherland R.~S., 2019, arXiv e-prints, arXiv: 191009730
\bibitem[K\"oppen \& Hensler(2005)]{koppen05} K\"oppen J., \& Hensler G., A\&A 434, 531 
\bibitem[Law et al.(2015)]{law15} Law D.~R., Yan R., Bershady M.~A. et al., 2015,  AJ, 150, 19
\bibitem[Liang et al.(2006)]{liang06} Liang Y.~C., Yin S.~Y., Hammer F. et al., 2006, ApJ, 652, 257
\bibitem[Maiolino \& Mannucci(2019)]{maiolino19} Maiolino, R., \& Mannucci, F. 2019, A\&ARv, 27, 3
\bibitem[Marino et al.(2013)]{marino13} Marino R.~A., Rosales-Ortega F.~F., S\'anchez S.~F. et al., 2013, A\&A, 559, 114
\bibitem[Matsuoka et al.(2018)]{matsuoka18} Matsuoka, K.,  Nagao T.,  Marconi  A.,  Maiolino  R.,  Mannucci  F., Cresci  G.,  Terao  K.,  Ikeda H., 2018, A\&A, 616L, 4  
\bibitem[Matsuoka et al.(2009)]{matsuoka09} Matsuoka K., Nagao T., Maiolino R., Marconi A., TaniguchiY., 2009, A\&A, 503, 721
\bibitem[Mignoli et al.(2019)]{mignoli19} Mignoli M., Feltre A., Bongiorno A. et al., 2019, A\&A, 626, 9
\bibitem[Miller et al.(2011)]{miller11} Miller B. P., Brandt W.~N., Schneider D.~P., Gibson R.~R., Steffen A.~T., Wu J., 2011, ApJ, 726, 20
\bibitem[M\'olla \& D\'{\i}az(2005)]{molla05} Moll\'a M., \& D\'{\i}az A.~I., 2005, MNRAS, 358, 521
\bibitem[Morales-Luis et al.(2014)]{moralesluis14} Morales-Luis A.~B, P\'erez-Monterp E., S\'anchez Almeida J., Mu\~nos-Tu\~non C., 2014, ApJ, 797, 81
\bibitem[Nagao et al.(2006a)]{nagao06a} Nagao, T., Maiolino, R., Marconi, A. 2006a, A\&A, 447, 863
\bibitem[Nagao et al.(2006b)]{nagao06b} Nagao, T., Maiolino, R., Marconi, A. 2006b, A\&A, 459, 85
\bibitem[Pagel et al.(1979)]{pagel79} Pagel B. E. J., Edmunds M.~G., Blackwell D.~E., Chun M.~S., Smith G., 1979, MNRAS, 189, 95
\bibitem[Peimbert \& Peimbert(2010)]{peimbert10} Peimbert A., \& Peimbert M., 2010, ApJ, 724, 791
\bibitem[Peimbert, Peimbert, \& Delgado-Inglada(2017)]{peimbert17} Peimbert, M., Peimbert A.,  Delgado-Inglada G., 2017, PASP, 129, 082001
\bibitem[P\'erez-Montero \& Contini(2009)]{enrique09} P\'erez-Montero E., \& Contini T., 2009, MNRAS, 398, 949
\bibitem[P\'erez-Montero(2014)]{enrique14} P\'erez-Montero E., 2014, MNRAS, 441, 2663
\bibitem[P\'erez-Montero(2017)]{enrique17} P\'erez-Montero E., PASP, 129, 043001
\bibitem[P\'erez-Montero et al.(2019)]{enrique19} P\'erez-Montero E., Dors O.~L., V\'{\i}lchez J.~M., Garc\'{\i}-Benito R. et al., 2019, 489, 2652
\bibitem[Penston et al.(1990)]{penston90} Penston M.~V., Robinson A., Alloin D. et al., 1990, A\&A, 236, 53
\bibitem[Pettini \& Pagel(2004)]{pettini04} Pettini M., \& Pagel B.~E.~J. 2004, MNRAS, 348, L59
\bibitem[Pilyugin(2003)]{pilyugin03} Pilyugin L.~S., 2003, A\&A, 399, 1003 
\bibitem[Pilyugin et al.(2004)]{pilyugin04} Pilyugin L.~S., V\'{\i}lchez J.~M., Contini T., 2004, A\&A, 425, 849
\bibitem[Pilyugin \& Grebel(2016)]{leonid16} Pilyugin L.~S., \& Grebel E.~K., 2016, MNRAS, 457, 3678 
\bibitem[Pilyugin \& Mattsson(2011)]{pilyugin11} Pilyugin L.~S., \& Mattsson L., 2011, MNRAS, 412, 1145
\bibitem[Pilyugin et al.(2012)]{leonid12} Pilyugin L.~S., Grebel E.~K., Mattsson L., 2012, MNRAS, 424, 2316
\bibitem[Queyrel et al.(2012)]{queyrel12} Queyrel J., Contini T., Kissler-Patig M., et al. 2012, A\&A, 539, A93
\bibitem[Riffel et al.(2018)]{riffel18} Riffel R.~A., Hekatelyne C., Freitas I.~C., 2018, PASA, 35, 40.
\bibitem[Riffel et al.(2017)]{riffel17} Riffel R. A., Storchi-Bergmann T., Riffel R. et al. 2017, MNRAS, 470, 992
\bibitem[Raimann et al.(2000)]{raimann00} Raimann D., Storchi-Bergmann T., Bica E., Melnick J.,  Schmitt H. 2000, MNRAS, 316, 559
\bibitem[Sanders et al.(2016)]{sanders16} Sanders R.~L., Shapley A.~E., Kriek M. et al., 2016, ApJ, 825, L23
\bibitem[Sanders et al.(2019)]{sanders19} Sanders R.~L. Shapley, A.~E., Reddy N.~A. et al., 2019, arXiv e-prints, arXiv:1907.00013
\bibitem[Robinson et al.(1987)]{robinson87} Robinson A., Binette L., Fosbury R.~A.E., Tadhunter C.~N., 1987, MNRAS, 227, 97
\bibitem[Rembold et al.(2017)]{sandro17} Rembold S.~B., Shimoia J.,  Storchi-Bergmann T. el al., 2017, MNRAS, 472, 4382
\bibitem[Stasi\'nska(2006)]{grazina06} Stasi\'nska G., 2006, A\&A, 454, 127L 
\bibitem[Smith (1975)]{smith75} Smith H.~E., 1975, ApJ, 199, 591
\bibitem[Storchi-Bergmann et al.(1998)]{thaisa98} Storchi-Bergmann T.,  Schmitt H.~R.,  Calzetti D., Kinney A.~L., 1998, AJ, 115, 909
\bibitem[Storchi-Bergmann et al.(1994)]{thaisa94} Storchi-Bergmann T, Calzetti D., Kinney A. L., 1994, ApJ, 429, 572
\bibitem[Thomas et al.(2019)]{thomas19} Thomas A.~D., Kewley L.~J., Dopita M.~A. et al, 2019, ApJ, 874, 100
\bibitem[van Zee et al.(1998)]{vanzee98} van Zee L., Salzer J.~J., Haynes M.~P., O'Donoghue A.~A., Balonek T.~J., 1998, AJ, 116, 2805 
\bibitem[Vaona et al.(2012)]{vaona12} Vaona L., Ciroi S., Di Mille F., Cracco V., La Mura G., Rafanelli P., 2012, MNRAS, 427, 1266
\bibitem[Veilleux \& Osterbrock(1987)]{veilleux87} Veilleux S., \& Osterbrock D.~E., 1987, ApJS, 63, 295
\bibitem[Vila-Costas \& Edmunds(1993)]{vilacostas93} Vila-Costas M.~B., \& Edmunds M.~G., 1992, MNRAS, 265, 199
\bibitem[Vila-Costas \& Edmunds(1992)]{vilacostas92} Vila-Costas M.~B., \& Edmunds M.~G., 1992, MNRAS, 259, 121
\bibitem[Vincenzo \& Kobayashi(2018)]{vincenzo18} Vincenzo F., \& Kobayashi C., 2018, MNRAS, 478, 155
\bibitem[Zhang et al.(2013)]{zhang13} Zhang Z.~T., Liang Y.~C., Hammer F., 2013, MNRAS, 430, 2605
\bibitem[Zaritsky et al.(1994)]{zaritsky94} Zaritsky D., Kennicutt R.~C., Huchra J.~P., 1994, ApJ, 420, 87 
\bibitem[Zinchenko et al.(2019)]{igor19} Zinchenko I.~A., Dors O.~L.; H\"agele G.~F., Cardaci M.~V., Krabbe A.~C., 2019, MNRAS, 483, 1901
\bibitem[Zhu et al.(2019)]{zhu19} Zhu S.~F., Brandt W.~N., Wu J., Garmire G.~P., Miller B.~P., 2019, MNRAS, 482, 2016
\bibitem[Xiao et al.(2012)]{xiao12} Xiao T.,  Wang T., Wang H. et al., 2012, MNRAS, 421, 486
\bibitem[Yin et al.(2007)]{yin07} Yin S.~Y., Liang Y.~C., Hammer F. et al.,  2007, A\&A, 462, 535
\bibitem[Yates et al.(2012)]{yates12} Yates R.~M., Kauffmann G., Guo Q., 2012, MNRAS, 422, 215
\bibitem[York et al.(2000)]{york00} York D. G. et al., 2000, ApJ, 120, 1579
\end{thebibliography}
\end{document}